
\input amstex

\documentstyle{amsppt}
\NoRunningHeads
\magnification=1100

\hsize5.5in
\vsize7.25in
\hfuzz=25pt
\vcorrection{.25in}

\TagsOnRight

\vskip.25in

\topmatter
\title 
Finding small inhomogeneities from scattering data.
\endtitle

\author A.G.Ramm
\endauthor

\affil Department of Mathematics, Kansas State University,
Manhattan, KS 66506-2602, USA\\
email: {\it ramm\@math.ksu.edu}
\endaffil

\thanks{ The author thanks DAAD for support}
\endthanks
\abstract A new method for finding small
inhomogeneities from surface scattering data is proposed and
mathematically justified. The method allows one 
to find small holes and cracks in metallic and other obejcts from
the observation of the acoustic field   scattered by the
objects.\endabstract

\endtopmatter

\document

\subhead 1. Introduction
\endsubhead  \vskip.2in

In many applications one is interested in finding small
inhomogeneities in a medium from the observation of the scattered
field, acoustic or electromagnetic, on the surface of the
medium.

We have two typical examples of such problems in mind.
The first one is in the area of material science and technology.
Suppose that a piece of metal or other material is given
and one wants to examine if it has small cavities (holes or
cracks ) inside. One irradiates the metal by acoustic waves
and observes on the surface of the metal the scattered field.
From these data one wants to determine: 

1) are there small cavities inside the metal?

2) if there are cavities, then where are they located
and what are their sizes?

Similar questions can be posed concerning localization not
only of the cavities, but any small in comparison with the
wavelength, inhomogeneities. Our methods allow one to
answer such questions.

As a second example, we mention the mammography problem.
Currently x-ray mammography is widely used as a method of
early diagnistics of breast cancer in women. However,
it is believed that the probability for a woman to get a new
cancer cell in her breast as a result of an x-ray mammography
test is rather high (about 35 percent). Therefore it is
quite important to introduce ultrasound mammography tests.
This is being done currently. A new cancer cells can be
considered as small inhomogeneities in the healthy breast tissue.
The problem is to localize them from the observation
on the surface of the breast of the scattered
acoustic field. 

The purpose of this short paper is to describe a new idea of
solving the problem of finding
inhomogeneities, small in comparison with the wavelength,
from the observation of the scattered acoustic or electromagnetic waves
on the surface of the medium.

For simplicity we present the basic ideas in the case of acoustic
wave scattering.
These ideas are based on the earlier results on wave scattering
theory by small bodies, presented in [1]-[5].
Our objective in solving the inverse scattering problem of
finding  small inhomogeneities from surface scattering data
are:

1) to develop a computationally simple and stable method
for a partial solution to the above inverse scattering problem.
The exact inversion procedures (see [3] and references therein)
are computationally difficult and unstable. In practice it
is often quite important, and sometimes sufficient for
practical purposes, to get a "partial inversion",
that is, to answer questions of the type we asked above:
given the scattering data, can one determine if these data
correspond to some small inhomogeneities inside the body?
If yes, where are these inhomogeneities located?
What are their intensities?
We define the notion of intensity $v_m$ of an inhomogeneity below
formula (1).

In section 2 the basic idea of our approach is described.
In section 3 its short justification is presented.
Some theoretical and numerical
results based on a version of the proposed approach
one can find in [5].

\subhead 2. Basic equations
\endsubhead \vskip.02in
Let the governing equation be 
$$ [\nabla^2+k^2+k^2v(x)]u=-\delta(x-y)\ \text{in}\  {\Bbb R}^3,
   \tag1 $$ 
where $u$ satisfies the radiation condition, $k=const>0$, and $v(x)$ is 
the inhomogeneity in the velocity profile.

Assume that  $\sup_{x\in \Bbb R^3}|v(x)|\leq c_0$, 
supp $v=U^M_{m=1} B_m(\tilde z_m,\rho_m)\subset {\Bbb R}^3_-
=\{x:x_{(3)}<0\}$, where $x_{(3)}$ denotes the third component
of vector $x$ in Cartesian coordinates,
 $B_m(\tilde z_m,\rho_m)$ is a ball, centered
at $\tilde z_m$ with radius $\rho_m$, $k\rho_m\ll 1$. 

Denote
$$\overline {v_m}:=\int_{B_m}v(x)dx.$$

\demo{Inverse Problem (IP)} Given $u(x,y,k)$ for all 
$x,y\in P, P=\{x:x_{(3)}=0\}$ and  a fixed $k>0$, find 
$\{\tilde z_m,\overline v_m\}, \, 1\leq m \leq M$.
\enddemo

In this paper we propose a numerical method for solving
the (IP).

To describe this method let us introduce the following notations:
$$ P:=\{x:x_{(3)}=0\}, \tag2 $$
$$ \{x_j,y_j\}:=\xi_j,\quad 1\leq j\leq J,\quad 
   x_j,y_j\in P,$$
\vskip-.2in
$$ \ \text{are the points at which the data}
   \  u(x_j,y_j,k)\ \text{are collected}, \tag3 $$
$$ k>0\ \text{is fixed}, \tag4 $$
$$ g(x,y,k):=\frac{\exp(ik|x-y|)}{4\pi|x-y|}, \tag5 $$
$$ G_j(z):=G(\xi_j,z):=g(x_j,z,k)g(y_j,z,k), \tag6 $$
$$ f_j:=\frac{u(x_j,y_j,k)-g(x_j,y_j,k)}{k^2}, \tag7 $$
$$ \Phi(z_1,\dots,z_m,\, v_1,\dots,v_m):=
   \sum^J_{j=1}\left| f_j-\sum^M_{m=1} G_j(z_m)v_m\right|^2.  
   \tag8 $$

The proposed method for solving the (IP) consists in finding the
global minimizer of function (8). This minimizer 
$(\tilde z_1,\dots,\tilde z_m,\, \tilde v_1,\dots,\tilde v_m)$
gives the estimates  of the
positions $\tilde z_m$ of the small inhomogeneities 
and their intensities 
$\overline v_m$. This is explained in more detail below formula (14).
Numerical realization of the proposed method, including a 
numerical procedure for estimating the 
number $M$ of small inhomogeneities from the
surface scattering data is described in [8].

Our approach with a suitable modification
is valid in the situation when
the Born approximation fails, for example, in
the case of scattering by delta-type inhomogeneities [9].

In this case the basic condition 
$Mk^2c_0\rho^2<<1\,\, (*) $ which guarantees
the applicability of the Born approximation is violated. 
Here
$\rho:=\max_{1\leq m \leq M}\rho_m$ 
 and $c_0$ was defined below formula (1).
We assume throughout that $M$ is not very large, 
between $1$ and $15$.

 In the scattering by a delta-type
inhomogeneity the assumption is $c_0\rho^3=const:=V$
as $\rho \to 0$, so that for any fixed $k>0$ one has
$k^2c_0\rho^2= k^2 V\rho^{-1}\to \infty$
as $\rho \to 0$,  and clearly condition (*) is violated.

In our notations this delta-type inhomogeneity is
of the form $k^2v(x)=k^2 \sum_{m=1}^M \overline{v_m} \delta(x-\tilde
z_m)$. 

The scattering theory by the delta-type
potentials (see [9]) requires some facts from
the theory of selfadjoint extensions of
symmetric operators in Hilbert spaces and in this short
paper we will not go into detail. 

\subhead 3. Justification of the proposed method \endsubhead
\vskip.2in

We start with an exact integral equation equivalent to equation
(1) with the radiation condition:
$$ u(x,y,k)=g(x,y,k)+k^2 \sum^M_{m=1}\int_{B_m}
   g(x,z,k)v(z)u(z,y,k)dz.\tag 9 $$
For small inhomogeneities the integral on the right-hand side of
(9) can be approximately written as
$$ \align
k^2 \int_{B_m} g(x,z,k)v(z)u(z,y,k)dz 
   &:=k^2 \int_{B_m}g(x,z,k)v(z)g(z,y,k)dz+ \varepsilon^2
\tag10 \\
   & =k^2 G(x,y,\overline z_m)
   \int_{B_m}vdz+\varepsilon^2 \\
   &=k^2 G(\xi,\overline z_m)
\overline v_m +\varepsilon^2,\quad 1\leq m\leq M, 
    \endalign $$
where $\varepsilon^2$ is defined by the first equation
in formula (10), it is the error due to replacing
$u$ under the sign of integral in (9) by $g$,
$\overline z_m$ is a point close to $\tilde z_m$.

One has $|u-g|=O(\frac {Mk^2c_0\rho^3}{d^2})$ if $x,y\in P,$
so the error term $\varepsilon^2$ in (10) equals to
$O(\frac {M^2k^4 c_0^2\rho^6}{d^3})$ if $x,y\in P.$

Therefore the function
$u(z,y,k)$ under the sign of the integral in (9) can be replaced
by $g(x,y,k)$ with a small
error provided that 
$$  c_0M\frac {k^2\rho^3}{d}\ll 1,\quad x,y\in P, \quad d\sim 1, 
\tag11 $$
where
$ \rho=\max_{1\leq m\leq M} \rho_m, \quad
   c_0:=\max_{x\in{\Bbb R}^3} |v(x)|,$
 $M$ is the number of inhomogeneities, $d$ is the
minimal distance from $B_m$, $m=1,2,....,M$ to the surface $P$,
 $d\sim 1$ means that
the length is measured in the units of length $d$.

If a sufficient condition for the validity of the Born
approximation holds, that is, 
$$
Mk^2c_0\rho^2:=\delta <<1,\tag12
$$
then 
$$O(\frac {M^2k^4 c_0^2\rho^6}{d^3})= 
O(\frac {\delta^2\rho^2}{d^3})<<1 \quad \text { if } d\sim 1.
$$
Note that $u$ in (9) has dimension $L^{-1},$ where
$L$ is the length.

If the Born approximation is not valid, for example,
if $c_0\rho^3=V\neq 0$ as $\rho \to 0$, which is
the case of scattering by delta-type inhomogeneities,
then the error term $\varepsilon^2$ in formula (10) can still be
negligible: in this case  
$\varepsilon^2=O(\frac {M^2 k^4 V^2}{d^3}),$
so $\varepsilon^2 <<1$ if $\frac {M^2 k^4 V^2}{d^3}<< 1$.

If one understands a sufficient condition
for the validity of the Born approximation
as the condition which guarantees the smallness of 
$\varepsilon^2$ for all $x,y\in \Bbb R^3$ then condition (12)
is such a condition. However, if one understands 
a sufficient condition
for the validity of the Born approximation
as the condition which guarantees the smallness of  
$\varepsilon^2$ for  $x,y$ running only through the
region where the scattered field is measured,
in our case when  $x,y\in P$, then a much weaker
condition (11) will suffice.

In the limit $\rho \to 0$ and $c_0\rho^3=V\neq 0$ 
 formula (10) takes the form (13) (see below).
This can be derived from 
[9, p.113]. Formula (1.1.33) in [9] shows that the resolvent
kernel of the Schr\"odinger operator with the delta-type 
potential supported on a finite set of points (in our case
on the set of points $\tilde z_1, ...., \tilde z_M$)
has the form 
$$u(x,y,k)=g(x,y,k)+k^2\sum_{m=1}^M 
{c_{mm'}}g(x,\tilde z_m){g(y,\tilde z_{m'})}, \tag13
$$
where $c_{mm'}$ are some constants. These constants are
determined by a selfadjoint realization
of the corresponding
Schr\"odinger operator with delta-type potential.
There is an $M^2-$parametric family of such
realizations (see [9] for more details).
 
Although in general the matrix $c_{mm'}$ is not diagonal,
under a practically reasonable assumption (11)
one can neglect the off-diagonal terms of the matrix  $c_{mm'}$
and then formula (13) reduces practically to the form (10)
with the term $\varepsilon^2$ neglected.

We have assumed in (10) that the point $\overline {z_m}$
exists such that $\int_{B_m}g(x,z,k)v(z)g(z,y,k)dz=
G(x,y,\overline {z_m})\overline {v_m}$. This is
an equation of the type of mean-value theorem.
However, such a theorem does not hold, in general,
for complex-valued functions. Therefore,  
if one wishes to have a
rigorous derivation, one has to
add to  the error term $\varepsilon^2$ in (10) the
error which comes from replacing of the 
integral $\int_{B_m}g(x,z,k)v(z)g(z,y,k)dz$  in (10) 
by the term $G(x,y,\overline {z_m})\overline {v_m}$.
The error of such an approximation can be easily estimated.
We do not give such an estimate, because the basic conclusion
that the error term is negligible compared with the main term
$k^2G(x,y,\overline{z_m})\overline {v_m}$ remains valid
under our basic assumption  $k\rho<<1$.
From (10) and (7) it follows that
$$ f_j\approx\sum^M_{m=1}G_j(\overline z_m)\overline v_m, \quad
   G_j(\overline z_m):=G(\xi_j,\overline z_m,k) \tag14 $$
Therefore, parameters $\tilde z_m$ and $\overline v_m$ can be
estimated by the least-squares method if one finds the global
minimum of the function (8):
$$ \Phi(z_1,\dots,z_M,\, v_1,\dots,v_M)=\min. \tag15 $$
Indeed, if one neglects the error of the approximation (10),
then the function (8) is a smooth function of several variables,
namely, of $z_1,z_2,....z_m, v_1, v_2,...v_m$,
 and the global
minimum of this function is zero and is attained at the actual
intensities $\overline v_1,\overline v_2,.....,
\overline v_m$ and at the values 
$z_i=\overline z_i, i=1,2,....m$.

This follows from the simple argument: if the error
of approximation is neglected, then the approximate
equality in (14) becomes an exact one. Therefore
$f_j-\sum_{m=1}^M G_j(\overline z_m)\overline v_m=0$,
so that function (8) equals to zero. Since
this function is non-negative by definition,
it follows that the values $\overline z_m$ and $\overline v_m$ are
global minimizers of the function (8).
Therefore we take the global minimizers of
function (8) as approximate values of the positions and
intensities of the small inhomogeneities.
 
In general we do not know that the global minimizer is unique.
For the case of one small inhomogeneity $(m=1)$ uniqueness of the
global minimizer is proved in [5] for all sufficiently small
$\rho_m$ for a problem with a different functional. The problem
considered in [5] is the (IP) with $M=1$, 
and the functional minimized in [5] is specific for one
inhomogeneity.

The scattering theory for small scatterers originated in the 
 classical works of Lord Rayleigh. It was developed in [1] and [2],
 where analytical formulas for the scattering matrix
 were derived for the acoustic and electromagnetic scattering
 problems. In [1]
and [3] inverse scattering problems for small bodies are considered.
Numerically an important ingredient of our approach is the 
solution of the global minimization problem (14). 
The theory of global minimization is developed extensively
and the literature of this subject is quite large.
We mention two recent papers [6] and [7] where the reader can find many
references.

\vskip.5in
\vfill\eject


\Refs 
\widestnumber\key{LSW}

\vskip.2in

\ref\no 1 \by A.G. Ramm
\book Iterative methods for calculating the
static fields and wave scattering by small bodies
\publ Springer Verlag \publaddr New York \yr 1982
\endref

\vskip.1in
\ref\no 2\bysame
\book Scattering by obstacles
\publ D. Reidel \publaddr Dordrecht
\yr 1986 \pages 1--442
\endref

\vskip.1in
\ref\no 3\bysame
\book Multidimensional inverse scattering problems
\publ Longman/Wiley \publaddr New York \yr 1992 \pages 1--385
\transl expanded Russian edition \publ MIR
\publaddr Moscow \yr 1994 \pages 1--496
\endref

\ref\no 4 \bysame \paper A method for finding small inhomogeneities from
surface data \jour Math. Sci. Research Hot-Line \vol 1 \issue 10
\yr 1997 \pages 40-42
\endref
\ref\no 5 \by A.I.Katsevich and A.G. Ramm 
\paper Approximate inverse geophysical scattering on a small body
\jour SIAM J. Appl. Math. \vol 56 \issue N1 \yr 1996 
\pages 192--218 
\endref
\ref\no 6 \by J.Barhen, V.Protopopescu and D.Reister
\paper TRUST: A deterministic algorithm for global optimization
\jour Science \vol 276 \yr 1997 \pages 1094-1097
\endref
\ref\no 7 \by J.Barhen and V.Protopopescu
\paper Generalized TRUST algorithms for global optimization
\inbook State of the art in global optimization
\eds C.Floudas and P.Pardalos
\publ Kluwer Acad.
\publaddr Boston
\yr 1996 \pages 163-180
\endref
\ref\no 8 \by S.Gutman and A.G.Ramm
\paper Application of the hybrid stochstic-deterministic minimization method to a
surface data inverse scattering problem (submitted)
\endref
\ref\no 9\by S. Albeverio, F. Gesztesy, R. Hoegh-Krohn, H. Holden
\book Solvable models in quantum mechanics
\publ Springer Verlag \publaddr New York \yr 1988 \pages 1--452
\endref
\endRefs
\enddocument